%% file: main.tex
\documentclass[sigplan, 9pt]{acmart} 
\usepackage{multirow}
\AtBeginDocument{%
  }

\begin{document}

\title{\textit{DefensiveDR}: Defending against Adversarial Patches using \\Dimensionality Reduction}

\author{%
Nandish Chattopadhyay$^{1}$, Amira Guesmi$^{1}$, Muhammad Abdullah Hanif$^1$, \\ Bassem Ouni$^2$, Muhammad Shafique$^1$ \\
$^1$ eBrain Lab, Division of Engineering, New York University (NYU) Abu Dhabi, UAE \\ $^2$ AI and Digital Science Research Center, Technology Innovation Institute (TII), Abu Dhabi, UAE\\
}
\renewcommand{\shortauthors}{Chattopadhyay et al.}
\begin{abstract}
Adversarial patch-based attacks have shown to be a major deterrent towards the reliable use of machine learning models. These attacks involve the strategic modification of localized patches or specific image areas to deceive trained machine learning models. In this paper, we propose \textit{DefensiveDR}, a practical mechanism using a dimensionality reduction technique to thwart such patch-based attacks. Our method involves projecting the sample images onto a lower-dimensional space while retaining essential information or variability for effective machine learning tasks. We perform this using two techniques, Singular Value Decomposition and t-Distributed Stochastic Neighbor Embedding. We experimentally tune the variability to be preserved for optimal performance as a hyper-parameter. This dimension reduction substantially mitigates adversarial perturbations, thereby enhancing the robustness of the given machine learning model. Our defense is model-agnostic and operates without assumptions about access to model decisions or model architectures, making it effective in both black-box and white-box settings. Furthermore, it maintains accuracy across various models and remains robust against several unseen patch-based attacks. The proposed defensive approach improves the accuracy from 38.8\% (without defense) to 66.2\% (with defense) when performing LaVAN and GoogleAp attacks, which supersedes that of the prominent state-of-the-art like LGS \cite{naseer2019local} (53.86\%) and Jujutsu \cite{Jujutsu} (60\%).
\end{abstract}





\maketitle


\input{intro}
\input{background}
\input{methodology}

\input{experiments}

\input{related}

\input{conclusion}


\bibliographystyle{ACM-Reference-Format}
\bibliography{bib}

\appendix

\end{document}

%% file: intro.tex
\section{Introduction}
Through adversarial attacks, an attacker is able to significantly disrupt the performance of a highly trained deep neural network (DNN) model by introducing adversarial perturbation to the test samples \cite{10268441}. One specific type of adversarial attack involves the insertion of localized patches into the test image, forcing the model to make errors in tasks such as image classification or object detection. As these attacks become more prevalent, so does the attempt to defend against them and provide robustness.

Like in the case of most security related problems such as the simple adversarial attacks, the ongoing struggle lies in devising robust defenses that are resistant to exploitation. This perpetual game between attackers and defenders results in mutual reinforcement, each side striving to outsmart the other and grow stronger. 
Specifically, concerning adversarial patch-based attacks, the prevailing state-of-the-art, as discussed in the related work (Section \ref{related}), heavily relies on heuristics for patch identification and subsequent neutralization attempts. The inherent challenge emerges when inaccuracies arise in detecting the presence or absence of patches, subsequently affecting the entire defense strategy. This complication not only adds computational overhead but also introduces a susceptibility to errors in the overall process \cite{naseer2019local, xiang2021patchguard, Jujutsu}.

\begin{figure}[htbp]
\centerline{\includegraphics[width=\columnwidth]{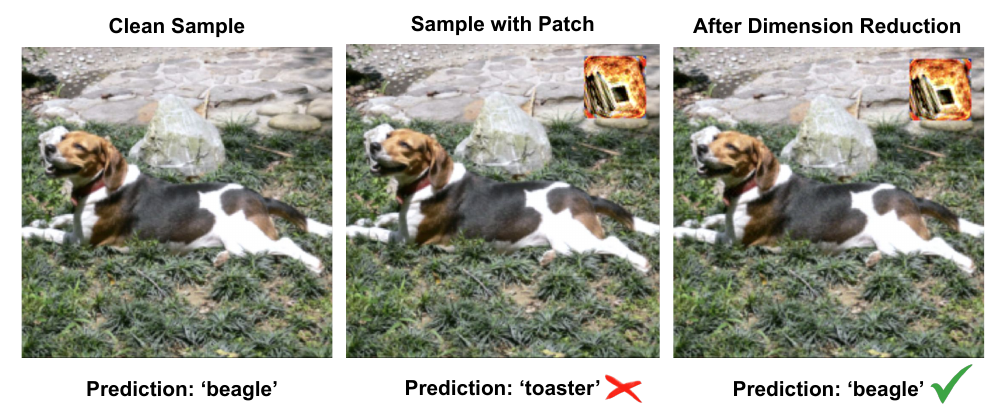}}
\caption{Dimensionality Reduction to thwart adversarial patch based attack.}
\label{fig:adv_patches_1}
\end{figure}

We take a step back from this and try to correlate the underlying geometry of the feature space in which the adversarial examples belong, along with the trained manifolds of the individual classes, to the success or failure of the attack. To that end, we make use of the understanding that adversarial attacks are more effective in the higher dimensional setting \cite{cod_1}, and that if the test samples (adversarial or otherwise) can be mapped to a lower dimensional setting, then this mapping can be done by statistical techniques \cite{freedman, tsne} that is able to segregate robust and non-robust features in the feature space, with some tuning parameter. This helps in making the model robust against the adversarial patches that the attacker inserts in the test samples, as shown in Figure \ref{fig:adv_patches_1}. 

\begin{figure}[!htbp]
\centerline{\includegraphics[width=\columnwidth]{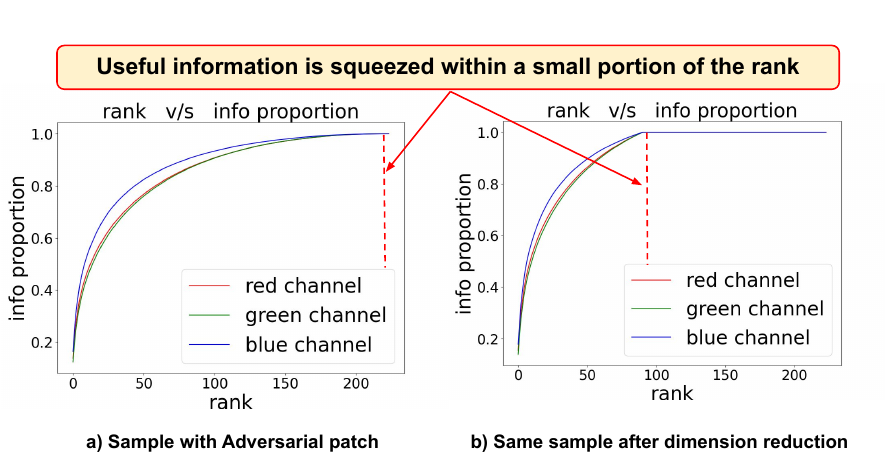}}
\caption{Using Dimensionality Reduction to separate adversarial noise from useful information necessary for machine learning task.}
\label{fig:adv_patches_2}
\end{figure}
The rationale behind employing Dimensionality Reduction is twofold. Firstly, reducing the samples to a lower-dimensional feature space significantly heightens the difficulty of executing successful adversarial attacks. Secondly, the process of dimensionality reduction itself possesses the capability to discern between robust and non-robust features, as illustrated in Figure \ref{fig:adv_patches_2}. It's noteworthy that, for a sample containing an adversarial patch, the information (captured by variability within the three matrices of Red, Blue, and Green color) is initially well-distributed across the rank of the matrix. However, during dimensionality reduction, the useful information within the same matrix becomes concentrated in a small portion of the rank, while the remaining portion contains noise. This intricate concept is elaborated in detail in Section \ref{method}. We harness this concept to formulate a practical defense mechanism against such attacks in the paper.

The \textbf{novel contributions} are summarized as follows:
\begin{itemize}
    \item We propose a novel and effective adversarial defense mechanism (\textit{DefensiveDR}) based on dimensionality reduction to defend against adversarial patches. In particular, 
    \textit{DefensiveDR} maps the input samples into a lower dimensional space in a way that it sufficiently segregates the robust and non-robust features to thwart the attack.
    
    \item The proposed defense is agnostic to the choice of neural architectures and can be deployed easily as a plug-in.
    \item We comprehensively analyse the effectiveness of our proposed defense for different DNNs and against two unseen patch-based attacks. Our technique achieves superior robust accuracy (67\%), and outperforms four existing defenses both certified and empirical in terms of robust accuracy: De-randomized smoothing (DS) \cite{levine2020randomized} (35.02\%), LGS \cite{naseer2019local} (53.86\%), PatchGuard \cite{xiang2021patchguard} (30.96\%), and Jujutsu \cite{Jujutsu} (60\%).  
    \item The proposed dimension reduction mechanism has negligible impact on model accuracy on clean images.
\end{itemize}

%% file: background.tex
\section{Background}
In this Section, we briefly touch upon the description of the attacks and the threat model, and the relationship of adversarial attacks and high dimensionality that we have exploited to implement the defense mechanism. 

\subsection{Patch-based Attacks}
Adversarial patches represent a specialized category of adversarial perturbations aimed at manipulating localized patches or specific regions within an image to deceive classification models. These attacks exploit the inherent vulnerability of models to localized alterations, with the ultimate goal of introducing subtle modifications that exert a substantial impact on the model's output. Leveraging the model's dependence on particular features or patterns, adversaries can craft patches designed to mislead the model into either misclassifying the image or perceiving it in a manner contrary to its intended interpretation.

\begin{figure}[htbp]
\centerline{\includegraphics[width=0.6\columnwidth]{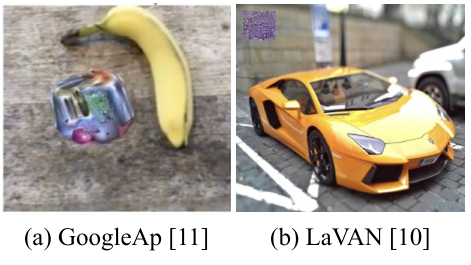}}
\caption{Examples of adversarial patches.}
\label{fig:adv_patches}
\end{figure}

Two well-known patch-based attacks are the following illustrated in Figure \ref{fig:adv_patches}:

\subsubsection{LaVAN \cite{lavan}}
LaVAN is a technique for generating localized and visible patches that can be applied across various images and locations. This approach involves training the patch iteratively by selecting a random image and placing it at a randomly chosen location. This iterative process makes sure that the model can capture the distinguishing features of the patch across a range of scenarios, thereby enhancing its ability to transfer and its overall effectiveness.

\subsubsection{GoogelAp \cite{googleap}}
GoogelAp offers a more practical form of attack for real-world scenarios compared to Lp-norm-based adversarial perturbations, which require object capture through a camera. This attack creates universal patches that can be applied anywhere. Additionally, the attack incorporates Expectation over Transformation (EOT) \cite{eot} to enhance the strength of the generated adversarial patch.

\subsection{Attack Formulation}

In the context of image classification, consider a deep learning-based image classifier represented as $f: X \rightarrow Y$, which is the mapping of an input image $x$ from the set of images $X$ to an output class with label $y$ from the set of labels $Y$. An adversarial example, denoted as $x^*$, is given by:

\begin{equation}
\label{eq:adv}
     \begin{array}{lll}
         x^* \in X, \quad f(x) = y, \quad f(x^*) = y^*, \quad y \neq y^*  \nonumber 
     \end{array}
\end{equation}

Here, $y^*$ is the targeted label, and $x^*$ is the adversarial example generated from the original input $x$. In the context of patch-based attacks, a portion of the image is replaced by the patch denoted as $P$.

Technically, the formulation of an adversarial example with a generated patch is expressed as:

\begin{equation}
    x^* = (1 - m_P) \odot x + m_P \odot P  \nonumber 
\end{equation}

Here, $\odot$ represents component-wise multiplication, $P$ is the adversarial patch, and $m_P$ is a mask matrix that constrains the shape, size, and pasting position of the patch. The value of the pasting area is set to 1, and 0 elsewhere.

To ensure that the patch $P$ is input-agnostic, it is trained over a variety of images. In the LaVAN approach \cite{lavan}, the patch is trained for a fixed location for each input $x \in X$. In the case of GoogleAp, the patch is trained to be applied in any random location. To further enhance the robustness of patch $P$ and make it physically realizable, GoogleAp \cite{googleap} uses a EOT framework \cite{eot}. EOT or Expectation over Transformation essentially uses various environmental transforms $T$ that can alter $x$ in various physical environments, such as translation, rotation, or lightness changes. Adversarial examples generated under these different transformations aim to remain robust, thus enhancing the overall effectiveness of the attack.

\subsection{Threat Model}
In our scenario, the attacker works in a white-box scenario where the attacker has complete information of the victim DNN, including its architecture and parameters. This is akin to other proposed defenses \cite{levine2020randomized, xiang2021patchguard, naseer2019local}. The attacker may also have knowledge about the presence of the dimension reduction based defense mechanism being in place, and this will not help them in circumventing the proposed defense mechanism. 
In this context, the attacker's strategy involves substituting a specific portion of the image with an adversarial patch. This patch is confined to a well-defined area within the image, and the attacker's primary objective is to consistently induce targeted errors in terms of misclassification across all input instances.

\subsection{Properties of High Dimensional spaces}
High dimensional spaces, like that of the feature space of the image samples in an image classification problem or the optimization loss landscape of the neural network, have some counter intuitive properties that help in the generation of adversarial examples.


In general, one assumes that any $1$-dimensional Gaussian distribution must have the highest mass near the mean, but mathematically, this does not hold good for high dimensions. According to the Gaussian Annulus Theorem~\cite{hopcroft}, for high values of $d$, in a Gaussian distribution of dimensions $d$, which has a variance of unity in each direction, for  $\beta \leq \sqrt{d}$, all but at most $3e^{-c \beta^2}$ portion of the probability distribution is concentrated in the small annulus $\sqrt{d} - \beta \leq \lvert x\lvert \leq \sqrt{d} + \beta$,  $c$ being a positive constant value \cite{norm_math}.


Specifically, in the case of high dimensional distributions, at least a $1-\frac{2}{c}e^{-c^2/2}$ proportion of the volume of it has $\lvert x_1 \lvert \leq \frac{c}{\sqrt{d-1}}$, for any $c \geq 1$ and $d \geq 3$. This means that for the best case scenario of the least surface area, which is that of an unit ball, the majority fraction of the data points lie near the periphery \cite{beyer}. 


Now, in any other arbitrary geometry, like that of the trained manifolds of the individual classes in an image classification problem, the surface area would certainly be higher, meaning that the majority of the data points will be even closer to the boundary that in the case of of the $d$-dimensional ball. Therefore, samples can be made to cross the hyperplane classifier separating the manifolds with relative ease and this is why adversarial attacks are easier in high dimensional feature spaces \cite{cod_1, googledimension}.

\subsection{Relationship with Adversarial Attacks}

As briefly mentioned here, the properties of behaviour of data samples belonging to the high dimensional spaces contribute significantly in the generation of adversarial examples \cite{dube}. This is particularly attributed to the fact that the samples being close to the decision boundary is easily shifted across to the erroneous side of the hyperplane classifier, resulting in it being an adversarial sample. This small perturbation in the targeted direction as found out from the gradients used in training the model, is imperceptible by human vision. Adversarial attacks are therefore possible to carry out with smaller perturbations in higher dimensional feature spaces \cite{cod_1}. 

%% file: methodology.tex
\section{Defense Mechanism and Implementation} 
\label{method}
In this section, we present the adversarial defense mechanism and describe the underlying techniques necessary for the defense pipeline. 

\subsection{Defense Pipeline}
\begin{figure}[htbp]
\centerline{\includegraphics[width=\columnwidth]{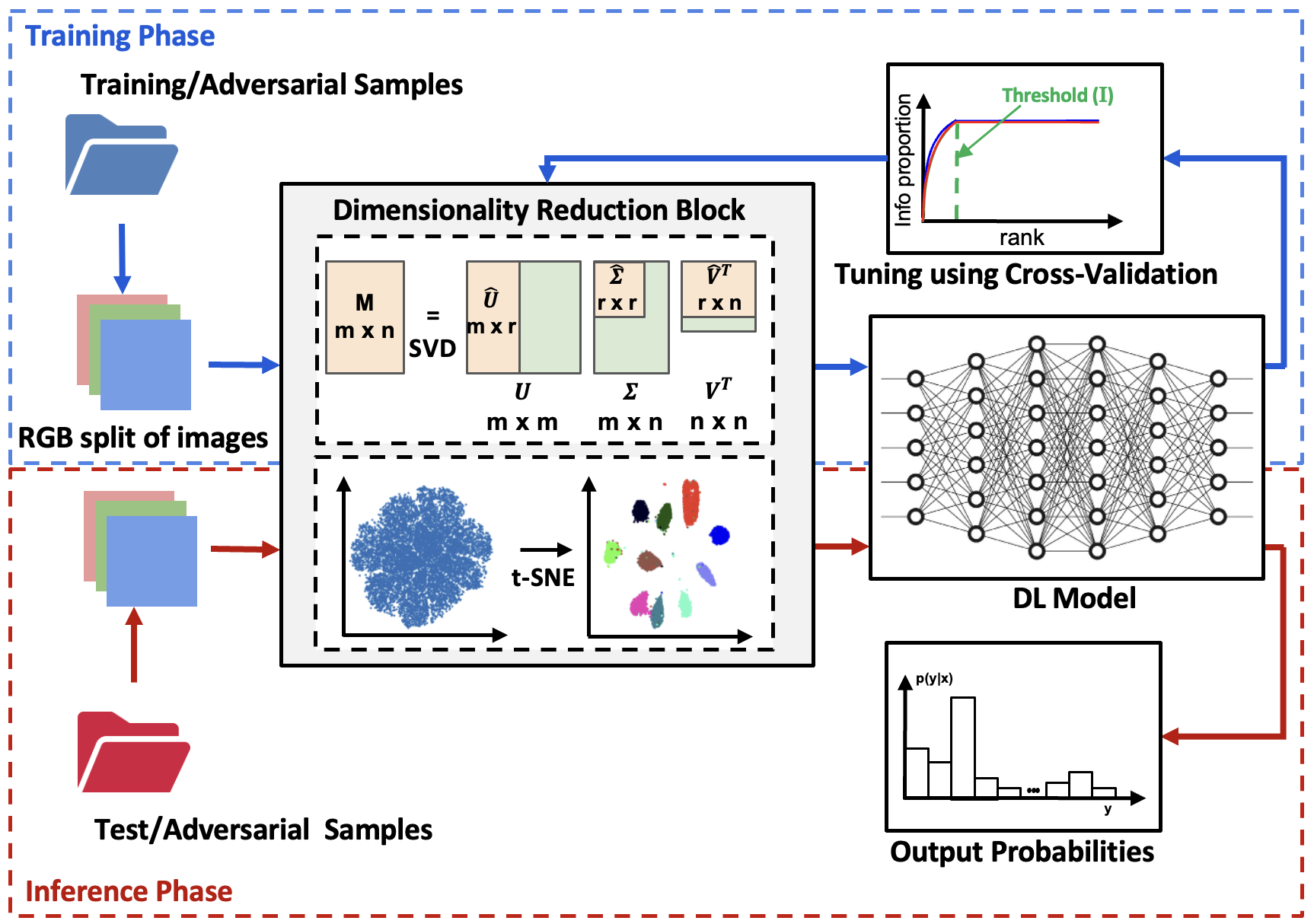}}
\caption{Overview of the \textit{DefensiveDR} Methodology.}
\label{fig:pipeline}
\end{figure}

Here, we explain how we integrate the dimensionality reduction block into the system and tune it for optimal performance. This is presented in a schematic form in Figure \ref{fig:pipeline}. Like any machine learning application, this has two phases, for training and inference. During the training phase, we use a mix of clean training samples and adversarial samples with the patches (in equal proportion) and apply dimension reduction on them, with a particular level of information content $I$. We iterate over different values of $I \in 99\%, \ldots, 90\%$ and check for which setting the performance of the model on adversarial samples is the highest whilst the drop in performance on the clean samples is less than $2\%$. This error rate is the level of tolerance we set, for incorporating robustness. The optimal value of $I$ is chosen and then, during the Inference phase, the same value of $I$ is used to map the test samples down to lower dimensions. The cross-validated tuning parameter helps us in sufficiently segregating the robust and the non-robust features during the mapping of the samples from the original dimension to a lower dimensional space.
The details of how the dimension reduction is done, and the motivation behind doing the same is explained hereafter. 

\subsection{Information vs Adversarial Noise}
Adversarial defenses that work in the feature space try to identify and isolate features, which are either necessary for the machine learning task and/or vulnerable to adversarial perturbation. Like any machine learning application, feature selection is of high importance for image classification as well and for this purpose, segregating robust features is necessary. There are many different works that try to achieve this using Feature Separation \cite{kim2023feature,features} wherein the prior models learn non-robust features from the dataset. The underlying principle of our approach is to correlate the robust features that the machine learning model makes use of, and the non-robust features that are vulnerable to adversarial attacks, which in this case is in the form of adversarial patches. As briefly outlined in the earlier section, this task becomes easier in lower dimensional spaces, and is difficult in the higher dimensional spaces owing to its properties and the behaviour of samples in such spaces. The mapping of the samples from the higher dimensional spaces to the lower dimensional spaces forces feature selection and separation of the robust features, which can be improved significantly using a relatively small effort in fine tuning the models. For this implementation, we have used two kinds of dimensionality reduction techniques, Singular Value Decomposition and t-Distributed Stochastic Neighbour Embedding.

\subsection{Dimensionality Reduction}
Dimensionality Reduction provides two benefits to thwart adversarial attacks. Firstly, adversarial attacks (both sparsely distributed and adversarial patches) are easier in higher dimensions than in lower dimensions because of the distribution of samples in the geometry of higher dimensional trained manifolds \cite{cod_1, dube}. Secondly, while mapping the samples from an higher dimensional space to a lower dimensional one, there are ways to select features that are critical for the machine learning task, while ignoring others that are vulnerable to adversarial attacks \cite{cod_2}. This fact is also facilitated by the imperceptibly property of the attacks, which means that the attacks like the adversarial patches in this case can not be mounted on the important features and are typically distributed among the non-robust features.

\subsubsection{Singular Value Decomposition (SVD)}
Singular Value Decomposition is a technique of dimensionality reduction that originates from linear algebra and may be very useful in the decomposition of a matrix, akin to the images that are used in our work. The extracted components are able to represent the variability or information contained in the matrices in forms that are more useful and interpretable \cite{freedman}. Essentially, this method lets us take a projection of the said matrix on to an orthonormal basis and the informational variability can then be expressed as a linear combination of its components.  

Considering an image to be an $m x n$-matrix $M$, one can use Singular Value Decomposition to factorize the same into $M = U \Sigma V^{T}$, where $U$ and $V$ are orthogonal matrices and $\Sigma = diag(\sigma_{1}, \ldots, \sigma_{r})$, where $r=min(m,n)$ such that $\sigma_{1} \geq \ldots \geq \sigma_{r} \geq 0$. In this case, the singular values are the $\sigma_{i}$s, and thereby we have the top $r$ columns of $V$ and $U$ being the right and left singular vectors respectively. 
This decomposition is used in the actual dimensionality reduction technique, wherein some components are preserved and some are discarded in what is called the Truncated SVD. 

For the matrix $M$ with a rank $r$, one can represent $M = U_{r} \Sigma_{r} V_{r}^{T}$ where $U_{r}$ and $V_{r}$ comprises of the left and right singular vectors. For any $k$, such that $(k <r)$, one can therefore obtain $M_{k}$, where $M_{k} = U_{k} \Sigma_{k} V^{T}_{k}$, which is lower rank approximation of the matrix. 
It may be noted here that the singular values are arranged in such a way that they are able to maintain the order in which the information is contained within each of the components and this is used for reducing dimensionality. In the proposed defense pipeline, the proportion of the singular values has been used as a surrogate of the fraction of information preserved when the image $M$ is projected to a lower dimension. The corresponding parameter for information preservation is denoted by $I$ and is tuned for every specific case. 


\subsubsection{t-SNE}
This algorithm provides a non-linear way of dimensionality reduction. In general, Stochastic Neighbour Embedding converts Euclidean distances between data samples at high-dimensions into conditional probabilities that are able to capture similarities \cite{tsne}. 
Let us consider that for a sample image (since we are working with image classification), a constituent vector $x_{j}$ and any other vector $x_{i}$, we have a measure of similarity expressed in their conditional probability $p_{j|i}$ such that these two vectors are neighbours, if neighbours are picked in the proportion of their probability density under some distribution assumption. The t-SNE algorithm uses a Student-t distribution for the similarity computation. Now, akin to the conditional probability $p_{j|i}$ in the high dimensional space, let us assume we have $q_{j|i}$ to be the conditional probabilities between the vectors in the low dimensional space $y_{j}$ and $y_{i}$. The mapping between them is established by minimizing the sum of the Kullback-Leibler divergences between the two distributions, that is the two conditional probabilities $p_{j|i}$ and $q_{j|i}$. In practice, the algorithm minimizes a single KL divergence $C$ between the two distributions. That is, the joint probability distributions $P$ and $Q$ in the high dimensional feature space and low dimensional feature spaces respectively. Therefore,
\begin{equation}
    C = KL(P||Q) = \sum_{i} \sum_{j} p_{ij} log \frac{p_{ij}}{q_{ij}} \nonumber
\end{equation}
For symmetric stochastic neighbour embedding, we have $p_{ii}$ and $q_{ij}$ equal to zero. Therefore, the upon mapping the samples $x$s in the high dimensional feature space to the corresponding samples $y$s in the low dimensional feature space, the pairwise similarities are:
\begin{equation}
    q_{ij} = \frac{exp (-||y_{i}-y_{j}||^{2})}{\sum_{k \neq l} exp (-||y_{k}-y_{l}||^{2})} \nonumber
\end{equation}
such that in the high dimensional space, the pairwise similarities are given by:
\begin{equation}
    p_{ij} = \frac{exp (-||x_{i}-x_{j}||^{2} / 2 \sigma^{2})}{\sum_{k \neq l} exp (-||x_{k}-x_{l}||^{2} / 2 \sigma^{2})} \nonumber
\end{equation}

In this case, we use the perplexity parameter to tune the functioning of t-SNE and vary the number of components as the tuning parameter $I$ to control the amount of information to be preserved upon dimensionality reduction.

%% file: experiments.tex
\section{Experimental Results}
In this section, we thoroughly assess the effectiveness of our proposed defense mechanism. 
\subsection{Experimental Setup}
\subsubsection{Datasets and Networks}
We conducted our defense evaluation on ImageNet \cite{imagenet} using three pretrained deep neural networks (DNNs) available in the TorchVision library: Resnet-50 \cite{he2015deep}, Resnet-152 \cite{he2015deep}, and VGG-19 \cite{simonyan2015deep}. These well-established models served as the basis for our assessment, ensuring a comprehensive evaluation across a range of DNN architectures and model complexities.

\subsubsection{Attack Setup}
The attacker's objective is to create adversarial patches that effectively deceive the victim deep learning-based classifier. We generate distinct patches in five different sizes: $38 \times 38$, $41 \times 41$, $44 \times 44$, $47 \times 47$, and $50 \times 50$. In the case of the LaVAN patch, the patch's location is fixed to the upper right corner of the image. On the other hand, for GoogleAp, the patch is placed randomly within the image. Each patch undergoes a training process comprising 100 epochs using a training dataset consisting of 1000 images. Subsequently, we assess the attack success rate on a separate test dataset. In the context of ImageNet, our evaluation employs a set of 10,000 images drawn from the validation dataset.

\subsubsection{Defense Setup}
The defense mechanism comprises of two phases, as shown in Section \ref{method}. In the Training phase, we use a $50-50\%$ proportion of samples of $1000$, from the ImageNet validation dataset to fine tune the model and also set the optimal working parameter $I$ for the dimension reduction block. We systematically experimented with different values of the Variability parameter, specifically considering settings $I \in 99\%, \ldots, 90\%$. Through empirical analysis, we identified the parameter value that achieves the optimal balance between robust accuracy, which measures the model's performance under adversarial conditions, and baseline accuracy, which reflects its performance on clean, unaltered data. In the Inference phase, the samples are passed through the dimension reduction block, followed by the actual model for classification. We have repeated this setup for the combination of models and attacks mentioned below.


\subsection{Evaluation of Defense Performance}
In our evaluation, we primarily focused on measuring the model's robust accuracy as the key metric for assessing the effectiveness of our defense technique. To illustrate the impact of our defense strategy, we initially generated adversarial patches using two distinct attack strategies, namely Lavan and GoogleAp. Subsequently, we reported the model's robust accuracy across different patch sizes, various models (Resnet-50, Resnet-152, and VGG-19), and different variability percentages. We conducted this assessment while utilizing two dimensionality reduction techniques, namely SVD and t-SNE, to provide a comprehensive analysis of our defense approach's performance under various scenarios and configurations.

As demonstrated in Tables \ref{tab1} and \ref{tab2}, our defense technique achieves a remarkable level of robust accuracy. The tables report the neural architecture, the clean accuracy of the neural network on the validation dataset, the drop in accuracy upon introducing the adversarial patch, and the impact on robustness by using dimension reduction. Specifically, for each of the two dimension reduction techniques, we report the performance of the model on samples containing the adversarial patches and samples without the patches.

\begin{table}[]
\caption{Robustness using dimensionality reduction for the GoogleAp attack \cite{googleap} on the Imagenet dataset}
\label{tab1}
\resizebox{0.49\textwidth}{!}{%
\begin{tabular}{l|l|l|l|l|ll|ll}
\hline
\multicolumn{1}{c|}{\multirow{2}{*}{\begin{tabular}[c]{@{}c@{}}Patch\\ Size\end{tabular}}} & \multirow{2}{*}{\begin{tabular}[c]{@{}l@{}}Model /\\ Neural\\ Network\end{tabular}} & \multirow{2}{*}{\begin{tabular}[c]{@{}l@{}}Clean \\ Acc\end{tabular}} & \multirow{2}{*}{\begin{tabular}[c]{@{}l@{}}Adv\\ Patch\\ Attack\end{tabular}} & \multirow{2}{*}{\begin{tabular}[c]{@{}l@{}}Info\\ \%\end{tabular}} & \multicolumn{2}{c|}{\begin{tabular}[c]{@{}c@{}}Dimension\\ Reduction:\\ SVD\end{tabular}} & \multicolumn{2}{c}{\begin{tabular}[c]{@{}c@{}}Dimension\\ Reduction:\\ t-SNE\end{tabular}} \\ \cline{6-9} 
\multicolumn{1}{c|}{} &  &  &  &  & \multicolumn{1}{l|}{\begin{tabular}[c]{@{}l@{}}Robust\\ (w/\\ patch)\end{tabular}} & \begin{tabular}[c]{@{}l@{}}Robust\\ (w/o\\ patch)\end{tabular} & \multicolumn{1}{l|}{\begin{tabular}[c]{@{}l@{}}Robust\\ (w/\\ patch)\end{tabular}} & \begin{tabular}[c]{@{}l@{}}Robust\\ (w/o\\ patch)\end{tabular} \\ \hline \hline 
\multirow{3}{*}{\begin{tabular}[c]{@{}l@{}}38\\ x\\ 38\end{tabular}} & ResNet152 & 81.2\% & 39.9\% & 95\% & \multicolumn{1}{l|}{66.5\%} & 78.1\% & \multicolumn{1}{l|}{\textbf{66.8\%}} & 78.5\% \\ \cline{2-9} 
 & ResNet50 & 78.4\% & 38.8\% & 95\% & \multicolumn{1}{l|}{\textbf{66.2\%}} & 76.2\% & \multicolumn{1}{l|}{65.9\%} & 76.7\% \\ \cline{2-9} 
 & VGG19 & 74.2\% & 39.1\% & 95\% & \multicolumn{1}{l|}{67.6\%} & 71.3\% & \multicolumn{1}{l|}{\textbf{68.1\%}} & 71.6\% \\ \hline
\multirow{3}{*}{\begin{tabular}[c]{@{}l@{}}41\\ x\\ 41\end{tabular}} & ResNet152 & 81.2\% & 21.4\% & 99\% & \multicolumn{1}{l|}{52.9\%} & 80.2\% & \multicolumn{1}{l|}{\textbf{53.1\%}} & 80.8\% \\ \cline{2-9} 
 & ResNet50 & 78.4\% & 21.1\% & 99\% & \multicolumn{1}{l|}{53.3\%} & 77.1\% & \multicolumn{1}{l|}{\textbf{53.5\%}} & 77.4\% \\ \cline{2-9} 
 & VGG19 & 74.2\% & 22.8\% & 99\% & \multicolumn{1}{l|}{53.8\%} & 73.6\% & \multicolumn{1}{l|}{\textbf{54.1\%}} & 73.9\% \\ \hline
\multirow{3}{*}{\begin{tabular}[c]{@{}l@{}}44\\ x\\ 44\end{tabular}} & ResNet152 & 81.2\% & 14.6\% & 90\% & \multicolumn{1}{l|}{\textbf{46.3\%}} & 77.8\% & \multicolumn{1}{l|}{45.9\%} & 78.6\% \\ \cline{2-9} 
 & ResNet50 & 78.4\% & 14.2\% & 90\% & \multicolumn{1}{l|}{46.6\%} & 74.4\% & \multicolumn{1}{l|}{\textbf{46.8\%}} & 74.2\% \\ \cline{2-9} 
 & VGG19 & 74.2\% & 15.8\% & 90\% & \multicolumn{1}{l|}{\textbf{45.9\%}} & 70.9\% & \multicolumn{1}{l|}{45.8\%} & 70.6\% \\ \hline
\multirow{3}{*}{\begin{tabular}[c]{@{}l@{}}47\\ x\\ 47\end{tabular}} & ResNet152 & 81.2\% & 9.3\% & 95\% & \multicolumn{1}{l|}{\textbf{36.9\%}} & 78.1\% & \multicolumn{1}{l|}{37.3\%} & 78.5\% \\ \cline{2-9} 
 & ResNet50 & 78.4\% & 9\% & 95\% & \multicolumn{1}{l|}{36.5\%} & 76.2\% & \multicolumn{1}{l|}{\textbf{36.9\%}} & 76.7\% \\ \cline{2-9} 
 & VGG19 & 74.2\% & 10.6\% & 95\% & \multicolumn{1}{l|}{35.9\%} & 71.3\% & \multicolumn{1}{l|}{\textbf{36.1\%}} & 71.6\% \\ \hline
\multirow{3}{*}{\begin{tabular}[c]{@{}l@{}}50\\ x\\ 50\end{tabular}} & ResNet152 & 81.2\% & 4.9\% & 95\% & \multicolumn{1}{l|}{23.9\%} & 78.1\% & \multicolumn{1}{l|}{\textbf{24.7\%}} & 78.5\% \\ \cline{2-9} 
 & ResNet50 & 78.4\% & 4.5\% & 95\% & \multicolumn{1}{l|}{24.5\%} & 76.2\% & \multicolumn{1}{l|}{\textbf{25.8\%}} & 76.7\% \\ \cline{2-9} 
 & VGG19 & 74.2\% & 3.8\% & 95\% & \multicolumn{1}{l|}{24.8\%} & 71.3\% & \multicolumn{1}{l|}{\textbf{25.9\%}} & 71.6\% \\ \hline
\end{tabular}%
}
\end{table}

\begin{table}[]
\caption{Robustness using dimensionality reduction for the LaVAN attack \cite{lavan} on the Imagenet dataset}
\label{tab2}
\resizebox{0.49\textwidth}{!}{%
\begin{tabular}{l|l|l|l|l|ll|ll}
\hline
\multicolumn{1}{c|}{\multirow{2}{*}{\begin{tabular}[c]{@{}c@{}}Patch\\ Size\end{tabular}}} & \multirow{2}{*}{\begin{tabular}[c]{@{}l@{}}Model /\\ Neural\\ Network\end{tabular}} & \multirow{2}{*}{\begin{tabular}[c]{@{}l@{}}Clean \\ Acc\end{tabular}} & \multirow{2}{*}{\begin{tabular}[c]{@{}l@{}}Adv\\ Patch\\ Attack\end{tabular}} & \multirow{2}{*}{\begin{tabular}[c]{@{}l@{}}Info\\ \%\end{tabular}} & \multicolumn{2}{c|}{\begin{tabular}[c]{@{}c@{}}Dimension\\ Reduction:\\ SVD\end{tabular}} & \multicolumn{2}{c}{\begin{tabular}[c]{@{}c@{}}Dimension\\ Reduction:\\ t-SNE\end{tabular}} \\ \cline{6-9} 
\multicolumn{1}{c|}{} &  &  &  &  & \multicolumn{1}{l|}{\begin{tabular}[c]{@{}l@{}}Robust\\ (w/\\ patch)\end{tabular}} & \begin{tabular}[c]{@{}l@{}}Robust\\ (w/o\\ patch)\end{tabular} & \multicolumn{1}{l|}{\begin{tabular}[c]{@{}l@{}}Robust\\ (w/\\ patch)\end{tabular}} & \begin{tabular}[c]{@{}l@{}}Robust\\ (w/o\\ patch)\end{tabular} \\ \hline \hline 
\multirow{3}{*}{\begin{tabular}[c]{@{}l@{}}38\\ x\\ 38\end{tabular}} & ResNet152 & 81.2\% & 10.1\% & 95\% & \multicolumn{1}{l|}{68.3\%} & 78.6\% & \multicolumn{1}{l|}{\textbf{68.9\%}} & 78.9\% \\ \cline{2-9} 
 & ResNet50 & 78.4\% & 10.2\% & 95\% & \multicolumn{1}{l|}{\textbf{68.8\%}} & 75.8\% & \multicolumn{1}{l|}{68.2\%} & 76.7\% \\ \cline{2-9} 
 & VGG19 & 74.2\% & 11.1\% & 95\% & \multicolumn{1}{l|}{67.1\%} & 71.9\% & \multicolumn{1}{l|}{\textbf{67.8\%}} & 71.9\% \\ \hline
\multirow{3}{*}{\begin{tabular}[c]{@{}l@{}}41\\ x\\ 41\end{tabular}} & ResNet152 & 81.2\% & 7.9\% & 95\% & \multicolumn{1}{l|}{\textbf{59.1\%}} & 78.6\% & \multicolumn{1}{l|}{58.9\%} & 78.9\% \\ \cline{2-9} 
 & ResNet50 & 78.4\% & 8.3\% & 95\% & \multicolumn{1}{l|}{59.4\%} & 75.8\% & \multicolumn{1}{l|}{\textbf{59.6\%}} & 76.7\% \\ \cline{2-9} 
 & VGG19 & 74.2\% & 8.1\% & 95\% & \multicolumn{1}{l|}{57.4\%} & 71.9\% & \multicolumn{1}{l|}{\textbf{57.8\%}} & 71.9\% \\ \hline
\multirow{3}{*}{\begin{tabular}[c]{@{}l@{}}44\\ x\\ 44\end{tabular}} & ResNet152 & 81.2\% & 4.9\% & 95\% & \multicolumn{1}{l|}{56.1\%} & 78.6\% & \multicolumn{1}{l|}{\textbf{56.5\%}} & 78.9\% \\ \cline{2-9} 
 & ResNet50 & 78.4\% & 4.8\% & 95\% & \multicolumn{1}{l|}{55.6\%} & 75.8\% & \multicolumn{1}{l|}{\textbf{55.8\%}} & 76.7\% \\ \cline{2-9} 
 & VGG19 & 74.2\% & 4.8\% & 95\% & \multicolumn{1}{l|}{55.2\%} & 71.9\% & \multicolumn{1}{l|}{\textbf{56.5\%}} & 71.9\% \\ \hline
\multirow{3}{*}{\begin{tabular}[c]{@{}l@{}}47\\ x\\ 47\end{tabular}} & ResNet152 & 81.2\% & 1.0\% & 90\% & \multicolumn{1}{l|}{24.4\%} & 78.1\% & \multicolumn{1}{l|}{\textbf{25.9\%}} & 78.5\% \\ \cline{2-9} 
 & ResNet50 & 78.4\% & 1.0\% & 90\% & \multicolumn{1}{l|}{24.6\%} & 74.7\% & \multicolumn{1}{l|}{\textbf{26.8\%}} & 73.9\% \\ \cline{2-9} 
 & VGG19 & 74.2\% & 1.1\% & 90\% & \multicolumn{1}{l|}{22.1\%} & 71.6\% & \multicolumn{1}{l|}{\textbf{24.9\%}} & 72.3\% \\ \hline
\multirow{3}{*}{\begin{tabular}[c]{@{}l@{}}50\\ x\\ 50\end{tabular}} & ResNet152 & 81.2\% & 1.9\% & 90\% & \multicolumn{1}{l|}{25.1\%} & 78.1\% & \multicolumn{1}{l|}{\textbf{28.1\%}} & 78.5\% \\ \cline{2-9} 
 & ResNet50 & 78.4\% & 2.0\% & 90\% & \multicolumn{1}{l|}{24.8\%} & 74.7\% & \multicolumn{1}{l|}{\textbf{25.9\%}} & 73.9\% \\ \cline{2-9} 
 & VGG19 & 74.2\% & 2.1\% & 90\% & \multicolumn{1}{l|}{23.9\%} & 71.6\% & \multicolumn{1}{l|}{\textbf{24.8\%}} & 72.3\% \\ \hline
\end{tabular}%
}
\end{table}

\begin{figure}[!htbp]
\centerline{\includegraphics[width=\columnwidth]{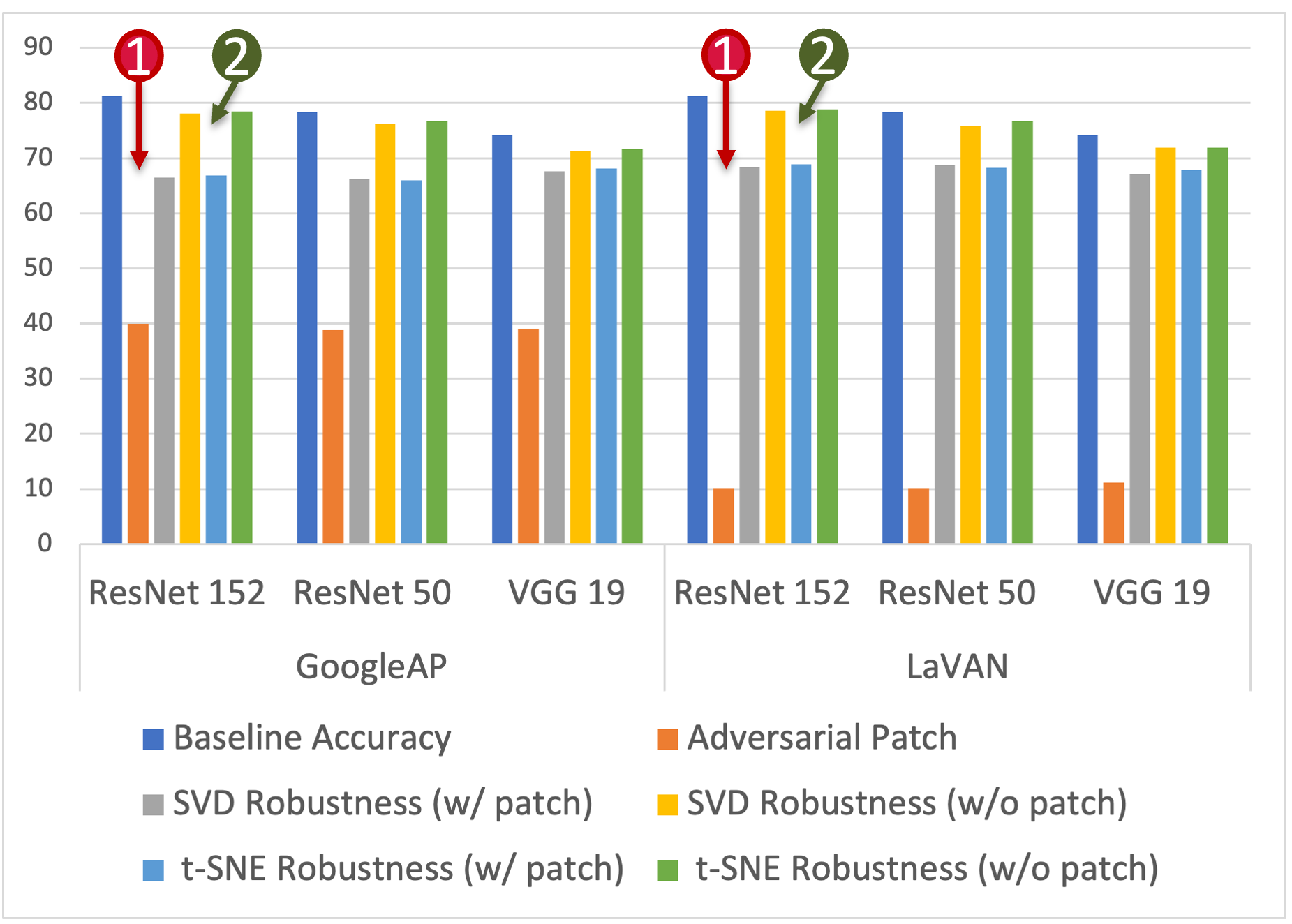}}
\caption{Effectiveness of Dimension Reduction to prevent adversarial patch based attacks, without compromising on functionality.}
\label{fig:plots}
\end{figure}

In Figure \ref{fig:plots}, we present the selected results form the aforementioned tables, across various neural architectures, for the patch size $38$ x $38$ where it is clear that the adversarial patch is significantly bringing down the accuracy of the model, compared to the baseline (see label 1). However, both the dimension reduction techniques bring up the accuracy when the adversarial patches are present, and also they do not disturb the performance of the model when adversarial patches are not present, thereby preserving functionality (see label 2).  

\subsection{Comparison with Related Techniques}
In our comparative analysis against four existing defense techniques, our approach demonstrated superior performance, achieving a robust accuracy of 66.2\% using the SVD technique for a patch size of $38 \times 38$ used to attack a Resnet50 model trained on ImageNet. This outperformed LGS with 53.86\%, DS with 35.02\%, PatchGuard with 30.96\%, and Jujutsu with 60\%. 

\begin{table}[]
    \centering
    \caption{Performance of our proposed defense compared to four state-of-the-art defenses against GoogleAp \cite{googleap} attack.}
    \begin{tabular}{c|c}
    \hline
     Defense   &  Robust Accuracy\\
     \hline
     LGS \cite{naseer2019local}    &  53.86\% \\
     \hline
     DS \cite{levine2020randomized}    & 35.02\% \\
     \hline
     PatchGuard \cite{xiang2021patchguard}  & 30.96\%\\
     \hline
     Jujutsu \cite{Jujutsu}  &  60\%\\
     \hline
     \textbf{Ours}   & \textbf{66.2\%} \\
     \hline
    \end{tabular}

    \label{tab:comaprison}
\end{table}

\subsection{Key Findings}
The key findings from our analysis are mentioned here:
\begin{itemize}
    \item The proposed dimensionality reduction technique for defending against adversarial attacks that involve insertion of adversarial patches work very well, providing at least $20\% - 25\%$ improvement in the accuracy of machine learning model at Inference. 
    \item This is consistent across different state-of-the-art models and the adversarial patches.
    \item The trade-off with the degradation of performance upon dimensionality reduction is safely negotiated, with the drop always within a $2-3\%$ range of the clean accuracy of the corresponding model without dimensionality reduction.
    \item The tuning parameter $I$ is of critical importance, which specifies how much information is to be preserved during dimensionality reduction to eliminate the adversarial noise. 
    \item Both the dimensionality reduction techniques work almost equally well, with the t-Distributed Stochastic Neighbour Embedding working slightly well than Singular Value Decomposition in most of the cases. 
    \item The strength of the patch based adversarial attack is strongly correlated with the patch size. It may also be noted that bigger patch sizes lead to lesser imperceptibility and therefore less practical attacks. Our proposed method is agnostic to the patch size and is able to maintain the performance benefits across different patch sizes and types of patches. 
\end{itemize}

%% file: related.tex
\section{Discussion and Prominent Related Work}
\label{related}

Defenses against adversarial patch-based attacks can be categorized into two main approaches: certified defenses and empirical defenses.

\textit{Certified defenses:} \textbf{De-randomized smoothing (DS)} \cite{levine2020randomized} introduces a certified defense technique by building a smoothed classifier by ensembling local predictions made on pixel patches. \textbf{PatchGuard}\cite{xiang2021patchguard} uses enforcing a small receptive field within deep neural networks (DNNs) and 
securing feature aggregation by masking out the regions with the highest sum of class evidence. 

\textit{Empirical defenses:} \textbf{Localized Gradient Smoothing (LGS)} \cite{naseer2019local} first normalizes gradient values and then uses a moving window to identify high-density regions based on certain thresholds.
\textbf{Jujutsu} \cite{Jujutsu} focuses on localizing adversarial patches and distinguishing them from benign samples. 

These defenses, while valuable, do come with certain limitations such as high false positive rates, 
poor detection rates etc. 
Another challenge is the ability to mitigate the adversarial impact effectively while still allowing deep neural networks (DNNs) to make correct inferences on the clean examples. 

%% file: conclusion.tex
\section{Conclusion}

We present a comprehensive investigation into defending against patch-based attacks in the context of deep learning models using dimensionality reduction, with \textit{DefensiveDR}. Through empirical evaluations, we have demonstrated the effectiveness of our defense approach in significantly enhancing model robustness. By optimizing key parameters and employing dimensionality reduction techniques (Singular Value Decomposition and t-SNE), we achieved impressive robust accuracy results on multiple patch based adversarial attacks and neural architectures, without compromising on the performance of the DNN model on benign samples. Additionally, \textit{DefensiveDR} also outperforms other defense techniques available in the literature in defending against unseen patch based adversarial attacks.